\documentstyle[preprint,prl,aps]{revtex}
\begin{document}
\preprint{\vbox{\hbox {November 1997} \hbox{IFP-745-UNC}  }}
%\twocolumn[\hsize\textwidth\columnwidth\hsize\csname@twocolumnfalse\endcsname

%\draft
\title{\bf Long-Lived Quarks?}
\author{\bf Paul H. Frampton$^{(a)}$ and 
Pham Quang Hung$^{(b)}$}
\address{(a)Department of Physics and Astronomy,}
\address{University of North Carolina, Chapel Hill, NC  27599-3255}
\address{(b)Department of Physics, University of Virginia,
Charlottesville, VA 22901.}
\maketitle
\begin{abstract}
Several lines of reasoning suggest that there might exist a 
non-sequential fourth generation
of heavy quarks having very small mixing
with light quarks and hence exceptionally long lifetime. 
\end{abstract}
\pacs{}
%\vskip0pc]
\newpage
The accurate measurement in 1989 of the width of the Z boson
\cite{Carter} showed that there exist precisely three 
neutrinos coupling as in the Standard Model (SM) and 
lighter than half the Z mass, $\sim 45$GeV. 

This 
led to the conventional wisdom that there 
are three and only three quark-lepton families. 
The discovery of the top quark\cite{CDF,D0} in 1995 was 
thus the final fermion of the SM. The only remaining 
particle is the Higgs boson expected to lie between 
$\sim 60$GeV and $\sim 600$GeV.  

This neat picture of 
just three quark-lepton families and a Higgs as the 
entire light spectrum of matter fields has great appeal. 
Depending on the mass of the Higgs\cite{Hung}, 
it could be the entire story up to the Planck or 
at least the GUT scale.  Nevertheless, apart from 
the unsatisfactory aspect that this picture does not 
explain why there are precisely three families there 
are two principal reasons for suspecting that it is 
incomplete and that there might exist more light particles: 

(1) The strong CP problem is unresolved. 
Although weak CP violation may be accommodated 
through the KM mechanism, solution of the strong CP issue 
is more satisfactorarily addressed by spontaneous CP 
violation in a model with two additional flavors of quark
\cite{aspon}.

(2) The three couplings of the SM fail to evolve to a common
unification point. Until recently this was thought\cite{ADF}
to offer support for low-energy supersymmetry, although
this has been questioned\cite{ADFFL} long ago. Very recently
one of us\cite{H} has pointed out that a fourth family
with a Dirac-mass neutrino, because it leads to
UV fixed points for the corresponding
Yukawa couplings, can lead to a satisfactory unification
at $\sim 3 \times 10^{15}$GeV.

These two quite different reasons lead us to the conclusion that
it is quite likely that the top is not the last flavor of quark
and that there likely
exists one further doublet $(U , D)$ which is
{\it either} vector-like ((1) above) {\it or}
chiral ((2) above).

The mass splitting of this extra doublet is severely constrained
by precision electroweak data, conveniently
parametrized by the S, T variables\cite{PT} -
the U variable is non-restrictive in this context.
For the non-chiral case, there is no contribution to
S and T at leading order unless the doublet has a mass
splitting leading to a contribution similar to
the first term in Eq(\ref{T}).

For the chiral case, there is a contribution to $T$ given by:
\begin{equation}
T = 
\frac{|\Delta M_Q^2|}{M_W^2}\frac{1}{4\pi sin^2\theta_W}  
+\frac{|\Delta M_L^2|}{M_W^2}\frac{1}{12\pi sin^2\theta_W}  
\label{T}
\end{equation}
where $M_Q$ is the heavy quark mass, $\Delta M_Q^2$ is the mass 
splitting in the quark doublet, and $\Delta M_L^2$ is
the corresponding mass splitting of the lepton doublet, assuming
no Majorana mass for $N_R$. Experimentally $T < 0.2$ so for
any $M_Q > 200$GeV, we deduce that the ratio of the U mass to the
D mass cannot exceed 1.1.
For the chiral case there is also a contribution to S given by,
for a complete chiral family:
\begin{equation}
S = \frac{2}{3\pi} = +0.21.    \label{S}
\end{equation}
which is only just compatible with precision data if $T < 0.2$.

This leads to our main point: the ratio of masses in
the fourth family is 1.1 or less, while
in the third family we have $m_t/m_b \sim 30$. In the
second, $m_c/m_s \sim 10$. This suggests heuristically that the fourth
family is very different, and hence likely to be
isolated from the first three families by tiny mixing angles. The question
then is how tiny this mixing might be and how long-lived these
heavy quarks can be. 
For example, let the mixing angle between the fourth family U quark and
b quark be $V_{Ub} = x$ and assume that $V_{Dt} =
V_{Ub}$. We shall particularly be interested in the following
two ranges: $ 10^{-5} < x < 10^{-3}$ and $x < 10^{-5}$. The first
range is the one considered by models such as the aspon model.
For $x < 10^{-5}$, we may assume an almost unbroken 3 + 1
structure under a horizontal family symmetry to isolate the fourth family
\cite{H2},
similar to the 2 + 1 family structure used in previous models\cite{kong}
which successfully accommodate the top quark mass.

Without specifying a particular model, we will examine the phenomenology
of heavy quarks of long lifetime corresponding to small
x in the quoted range. We will concentrate, in particular, on facilities
such as hadron colliders (Tevatron, LHC), where one has the best chance of
finding these objects. We will first estimate the production cross section
for such heavy quarks. We then discuss various signatures. 

For the Tevatron, the production cross section can be easily estimated since
it is similar to the one used for the top quark. The process $p \bar{p}
\rightarrow Q \bar{Q} + X$, where $Q$ is a heavy quark, can proceed through
$q \bar{q} \rightarrow Q \bar{Q}$ and $ g g \rightarrow Q \bar{Q}$. At the
Tevatron, the $q \bar{q}$ process dominates over the $gg$ one (roughly
90\% to 10\%). For example, if $m_Q = 180$ GeV, the cross section is
$\sim 4$ pb for $\sqrt{s} = 1.8$ TeV and $\sim 5.5$ pb for $\sqrt{s} =2$ TeV.
Also, for $\sqrt{s} = 2$ TeV, a heavy quark with mass $\sim 230$ GeV can have
a non-negligible cross section of $\sim 1.5$ pb. Of course, the cross section
increases tremendously at the LHC, by roughly three orders of magnitude.
Since the up and down heavy quarks (U and D) are supposed to be fairly degenerate,
their production cross sections are practically the same: we will have an equal
number of $U \bar{U}$ and $D \bar{D}$. Their signatures, however, are very 
different.

We shall assume that $m_U > m_D$.
Therefore both U and D can in principle have the following decay modes:
$U \rightarrow (D$ or q$) + (l^{+} \nu, q_1 \bar{q}_2)$,
$U \rightarrow q + W$, $D \rightarrow (t$ or $q) + (l^{-} \bar{\nu}, q_1 \bar{q}_2)$,
and $D \rightarrow (t$ or $q) + W$. Here $t$, $q$ and $l$ denote the top quark,
the light quark and the light lepton respectively. In those decays, one
has, in principle, to distinguish the chiral case from the vector-like one.
If $(U,D)$ is a chiral doublet, all of the above decays will be of the
V-A nature. If it is a vector-like doublet, the process
$U \rightarrow (D$ or q$) + (l^{+} \nu, q_1 \bar{q}_2)$ will contain both
V-A and V+A in the heavy quark current. However, because the light quarks
are chiral, it turns out that the decay rate is practically the same as if
$(U,D)$ were chiral. This is because the rate is proportional
to $(G_L^2 + G_R^2 )g_L^2 I_1 + G_L G_R g_L^2 I_2$ and that $I_2 \approx
-I_1$ and $G_L = G_R = g_L$ ($I_{1,2}$ are phase space integrals). The pure
V-A case will correspond to $G_R =0$ and one can see that the two rates
are the same.
For all other decays involving the transition between
a heavy and light quark, it will be pure V-A. Therefore, we shall only
list formulas related to the pure V-A cases.

The three-body
and two-body decay widths, $\Gamma_{3}^Q$ and $\Gamma_{2}^{Q}$ are given by
\begin{mathletters}
\begin{equation}
\Gamma_{3}^{Q_1} = 12 |V_{Q_1 Q_2}|^{2} \frac{ G_F(m_{Q_1})^{5}}{16 \pi^3} 
I_{3}(m_{Q_2}/m_{Q_1}, m_W /m_{Q_1}), \label{G3}
\end{equation}
\begin{equation}
\Gamma_{2}^{Q_1} = \frac{G_F (m_{Q_1})^{3}}{8 \pi \sqrt{2}} |V_{Q_1 Q_2}|^2 
I_{2} (m_{Q_2}/m_{Q_1}, m_W/m_{Q_1}), \label{G2}
\end{equation}
\end{mathletters}
\noindent
where $I_{3}$ and $I_{2}$ are well-known phase space factors\cite{BP}. Also, $V_{Q_1 Q_2}$
denotes the mixing between the two quarks, $Q_1$ and $Q_2$. For instance, we
may assume that $V_{UD} \approx 1$ and $|V_{Ub}| \approx |V_{Dt}| = x$, where $x$
is to be estimated. We shall start with the decay of the D quark first since
it will set the range of the mixing parameter $x$ where one can consider at least
one of the two heavy quarks to be long-lived. We then discuss the characteristic
signatures for such long-lived quarks.

The current accessible but unexplored decay length for a long-lived heavy quark 
to be detected is between $100 \mu m$ and $1 m$, a range on which we shall focus. 
(It should be noted that decay lengths less
than $100 \mu m$ and greater than $1 m$ are accessible as well with the latter
being explored. Also decay lengths of the order of a few tens of cm might
be hard to detect.)
Since $c \tau \sim 2 \times 10^{-10} GeV/
\Gamma (GeV) \mu m$, this would correspond to a width, $\Gamma$, between
$10^{-12}$ GeV and $10^{-16}$ GeV. (For comparison, the top quark width
is approximately 1.6 GeV.) 
 
We shall assume that $m_D \geq 175$ GeV.

For the special case of $m_D \approx 175$ GeV, we only have
$D \rightarrow (c, u) + W$. The decay width will be $\Gamma_{2}^{D} 
\approx 1.6 |V_{D(c,u)}|^2$ GeV. This would correspond to a lifetime
$\tau \approx 4 \times 10^{-25} |V_{D(c,u)}|^{-2}$ s, with
$c \tau \approx 1.2 \times 10^{-10} |V_{D(c,u)}|^{-2} \mu m$. The decay length
is given by $\beta \gamma c \tau$, with $\beta \gamma$ being typically of
order unity. The present experimental resolution is $\geq$ 100 $\mu m$. For a
175 GeV D quark to be detected, the mixing $|V_{D(c,u)}| < 10^{-6}$. In the most
simple minded scheme for the quark mass matrix, one might expect
$|V_{Dc}|$ to be at most $x^2$ if $|V_{Dt}| = x$ which would imply that
$x \leq 10^{-3}$. This, however, is highly model-dependent. It is best to
search for the direct decay of D into the top quark.

When $m_D > m_t$, D can decay into $t$ via the three and/or two body
processes, depending on its mass. For $m_D$ between $\sim 177$ GeV and
$m_t + m_W \sim 256$ GeV, the decay into $t$ is exclusively three-body.
However, because one can also have $D \rightarrow c + W$, the total
rate highly depends on how D mixes with $t$ and $c$. It is
straightforward to compute the width for 
$D \rightarrow t + (l^{+} \nu, q_1 \bar{q}_2)$ and compare that
with $D \rightarrow c + W$. To do this, let us put $|V_{Dt}| = x$
and $|V_{Dc}| = x^n$ where we shall assume that $n \geq 2$.

Let us first assume $x \leq 10^{-3}$. The detectability of the D quark
in the mass range from 177 to 190 GeV will depend on what $|V_{Dc}|$
might be. It goes as follows. Let us first assume $|V_{Dc}| = x^2$.
Let us also distinguish two classes of models: one with $x < 10^{-5}$
and one with $10^{-5} < x < 10^{-3}$ (the aspon model is one example of
the latter case.)

In Table 1, we show results for $\Gamma_{3}^{D} (D \rightarrow t) /x^2$,
$\Gamma_{2}^{D} (D \rightarrow t) /x^2$ and
$\Gamma_{2}^{D} (D \rightarrow c) /|V_{Dc}|^2$, as functions of $m_D$.
We see the following features.
I) For the first class of models with very small mixing,
$x < 10^{-5}$, it is straigthforward to see that one {\em always}
has $\Gamma_{3}^{D} (D \rightarrow t)  > \Gamma_{2}^{D} (D \rightarrow c)$.
Furthermore, one can also see that the decay of D will be detected
(within the range of $1 \mu m$ to $1 m$) only when $m_D > 190$ GeV (
corresponding to $\Gamma_{3} \sim 2 \times 10^{-16}$ for $x \sim 10^{-5}$).
The D quark with mass less than 190 GeV will just escape the detector
in this scenario. As the D quark mass increases beyond 190 GeV, the
mixing between D and $t$ will have to be smaller if one were to observe the D 
decay. For example, when $m_D = 310$ GeV, 
$\Gamma_{3}^{D} (D \rightarrow t)/x^2 \sim \Gamma_{2}^{D} (D \rightarrow t)/x^2
\sim 3$ GeV. This would require $x< 10^{-6}$. In any case, the
characteristic signature for this scenario is that D will
predominantly decay into the top quark!
II) For the second scenario with ``larger'' mixing (such as the aspon model),
namely $10^{-5} < x < 10^{-3}$, there is always an $x$ (e.g. $x \sim 10^{-3}$) 
where $\Gamma_{2}^{D} (D \rightarrow c)$ dominates over
$\Gamma_{3}^{D} (D \rightarrow t)$ for the range $m_D = 177-190$ GeV. So in
this range, it is possible that D will decay dominantly into the $c$ quark!
Since $\Gamma_{2}^{D} (D \rightarrow c)\propto x^4$, this decay is observable.
As $m_D > 190$ GeV, $\Gamma_{3}^{D} (D \rightarrow t)$ will dominate and
the decay is predominantly into $t$, just like the previous case.
However, as one can see from Table 1, this scenario is hard to observe
(i.e. the decay length is shorter than $1 \mu m$) for
$m_D \geq 250$ GeV ($\Gamma > 10^{-12}$ GeV). More sophisticated silicon
pixel detectors will be needed for such a case.

If $|V_{Dc}|= x^3$, $\Gamma_{3}^{D} (D \rightarrow t)$ will always be
greater than $\Gamma_{2}^{D} (D \rightarrow c)$ which means that
D will decay predominantly into $t$. The statements made above for
the first scenario with $x < 10^{-5}$ still hold in this case.
For the second scenario with $10^{-5} < x < 10^{-3}$, the ``detectability''
of the D quark is still possible as long as $m_D < 250$ GeV. However, in
contrast with the above discussion, the predominant decay mode is into the
top quark instead of the charm quark.

We now turn to the decay of the U quark. 
Since $m_U/m_D < 1.1$ ($\rho$ parameter
constraint), $m_U - m_D < 0.091 m_U < m_W$ unless $m_U > 890$ GeV, a strong
coupling scenario {\em not} considered in this paper. The U quark decays into
a D via a virtual W, namely $U \rightarrow D + (l^{+} \nu, q_1 \bar{q}_2)$,
where $l$ and $q$ are the light leptons and quarks. The U quark can also
decay into a light quark and a real W, namely
$U \rightarrow q + W$ where $q= b, s, d$ and with $U \rightarrow b$ assumed to be the
dominant transition.
Which decay mode of the U is dominant over
the other depends on how degenerate U is with D and on how small U mixes
with the $b$ quark. 
The results are shown in Table 2 where we list $\Gamma_{3}^{U}$ as a function
of the ratio $m_D/m_U$. As for $\Gamma_{2}^{U} (U \rightarrow b)$, the estimate
is straightforward. We obtain $\Gamma_{2}^{U} / x^2 \sim 1.75 - 4.7$ GeV
for $m_U = 180 - 250$ GeV. 

For the first scenario with $x < 10^{-5}$, we obtain
$\Gamma_{2}^{U} (U \rightarrow b) < (1.75 - 4.7) \times 10^{-10}$ GeV.
For the second scenario with $10^{-5} < x < 10^{-3}$, we obtain
$(1.75 - 4.7) \times 10^{-10} < \Gamma_{2}^{U} (U \rightarrow b) (GeV)
< (1.75 - 4.7) \times 10^{-6}$. These are to be compared with the results listed
in Table 2.

Unless the U and D quarks are very degenerate i.e. $m_D / m_U > 0.98$, the decay mode
$U \rightarrow D + (l^{+} \nu, q_1 \bar{q}_2)$ dominates in the first scenario
($x < 10^{-5}$). A look at Table 2 reveals that the U decay length is
{\em much less than} $1 \mu m$. The signals can be quite characteristic:
there is a primary decay of the U quark near the beam followed some $100 \mu m$
or so later by the secondary decay of the D quark. One might see two jets or
a charged lepton originating from near the colliding region followed by
three hadronic jets or one jet plus one charged lepton. The reconstruction of 
the event, if possible, might reveal the decay of a short-lived quark (the U
quark) into a long-lived quark (the D quark). There would approximately
an equal number $D \bar{D}$ produced and hence there might be events
where only the decay vertex of the D is seen.

For the second scenario with $10^{-5} < x < 10^{-3}$, we can see that,
if $m_D / m_U \geq 0.97$, $\Gamma_{2}^{U}$ dominates and U will decay
principally into $b$. One would not see the type of events with one
primary vertex separated by a hundred microns or so from the secondary one
as we have discussed above. For $m_D / m_U \leq 0.95$, the situation is
similar to the one encountered in the first scenario.

Thus far, we have assumed $m_U > m_D$ as in the second and third families.
We must not exclude the possibility that $m_D > m_U$ (as in the first family!).
The analysis we have given goes through {\it mutatis mutandis}, exchanging
$t \rightarrow b$, etc. A principal difference is that we may consider lighter
long-lived U quarks ({\it e.g.} $m_U < m_t$) than we did D quarks.

We thank A.T. Goshaw and David Stuart for a discussion of the Tevatron detectors. This
work was supported in part by the U.S. Department of Energy under 
Grant No. DE-FG05-85ER-40219 (PHF) and under Grant No. DE-A505-
89ER-40518 (PQH).

\begin{table}
\caption{The values of $\Gamma_{3}^{D} (D \rightarrow t + (l \nu, q \bar{q})/ x^{2}$,
$\Gamma_{2}^{D} (D \rightarrow c +W)/|V_{Dc}|^2$, and 
$\Gamma_{2}^{D} (D \rightarrow t +W)/x^2$ as functions of $m_D$}
% \label{}
\begin{tabular}{lccccccccc}
$m_D (GeV)$ & 177 & 180 & 190 & 200 & 220 & 250 & 270 & 290 & 310\\ \tableline
$\Gamma_{3}^{D}/ x^{2} (GeV)$ & $4.3 \times 10^{-11}$ & $1.1 \times
10^{-8}$ & $2 \times 10^{-6}$ & $2.7 \times 10^{-4}$ &
$3.8 \times 10^{-3}$ & $2.5 \times 10^{-2}$ &
2.8 & 3.1 & 2.8\\ \tableline
$\Gamma_{2}^{D} /|V_{Dc}|^2 (GeV)$ & 1.66 & 1.74 & 2.05 & 2.4 & 3.2
& 4.7 & 5.9 & 7.3 & 8.9 \\ \tableline
$\Gamma_{2}^{D} /x^2 (GeV)$ & 0 & 0 & 0 & 0 & 0 & 0 & 0.86 & 1.77
& 2.86 
\end{tabular}
\end{table}
\begin{table}
\caption{The width $\Gamma_{3}^{U} (U \rightarrow D + (l \nu, q \bar{q}))$ as a 
function of $m_D/m_U$}
%\label{}
\begin{tabular}{lccccc}
$m_D/m_U$ & 0.91 & 0.93 & 0.95 & 0.97 & 0.98 \\ \tableline
$\Gamma_{3}^{U}$ & $5.2 \times 10^{-5}$ & $1.5 \times 10^{-5}$ &
$2.8 \times 10^{-6}$ & $ 2.3 \times 10^{-7}$ & $3 \times 10^{-8}$
\end{tabular}
\end{table}


\begin{thebibliography}{99}
\bibitem{Carter}
J. R. Carter, {\it Precision Tests of the Standard Model at LEP}, in
Proc. LP-HEP 91, Editors: S. Hagerty, K. Potter and E. Quercigh,
World Scientific, Singapore (1992).
\bibitem{CDF}
F. Abe, {\it et al}. (CDF Collaboration), Phys. Rev. Lett. {\bf 74,} 2626 (1995).
\bibitem{D0}
S. Abachi, {\it et al}. (DO Collaboration), Phys. Rev. Lett. {\bf 74,} 2632 (1995).
\bibitem{Hung}
P.Q. Hung and G. Isidori, Phys. Lett. {\bf B402,} 122 (1997)
\bibitem{aspon}
P.H. Frampton and T.W. Kephart, Phys. Rev. Lett. {\bf 66,} 1666 (1991).\\
P.H. Frampton and D. Ng, Phys. Rev. {\bf D43,} 3034 (1991).
\bibitem{ADF}
U. Amaldi, W. de Boer and H. Furstenau, Phys. Lett. {\bf B260,} 240 (1991).
\bibitem{ADFFL}
U. Amaldi, W. de Boer, P.H. Frampton, H. Furstenau and J.T. Liu, Phys. Lett. {\bf B281,} 374 (1992).
\bibitem{H}
P.Q. Hung, {\it hep-ph/9710297}
\bibitem{PT}
M.E. Peskin and T. Takeuchi, Phys. Rev. Lett. {\bf 65,} 964 (1990); Phys. Rev. {\bf D46,} 381 (1992).
\bibitem{H2}
P. Q. Hung, Z. Phys. {\bf C74}, 561 (1997).
\bibitem{kong}
P.H. Frampton and O.C.W. Kong, Phys. Rev. {\bf D53,} R2293 (1996) and
references therein.
\bibitem{BP}
V. Barger and R.J.N. Phillips, Collider Physics, Updated Edition,
Addison-Wesley (1997).
\end{thebibliography}
\end{document}